\documentclass[aps,prl,twocolumn]{revtex4}

\usepackage{graphicx}
\newcommand{\etal}{\mbox{\it et al.}}

\begin{document}

\noindent {\bf Creffield and Sols Reply:}
Benenti \etal \cite{liars} assert that in our Letter \cite{prl}
we claimed that the ratchet current we observed for time-symmetric 
driving would persist indefinitely. Their assertion is
false. In our Letter we clearly indicated
that, in general, the ratchet currents would be transient,
and indeed wrote that we estimated them only to be
``stable over time scales ... of the order of 50 driving periods''. 
Unfortunately
Benenti \etal \ appear not to have read our paper with sufficient care
to have noted our discussion of this point, since we did not claim, 
or even imply, that the ratchet currents would be of infinite duration. 

To arrive at our estimate for the stability of the current,
we used a technique developed in
Ref. \onlinecite{shep} to estimate the Ehrenfest time 
of the system. 
In our study we considered a completely coherent time evolution, and
accordingly the current is given by a {\em coherent} sum
\begin{equation}
I(t) = \sum_{m,n} c_n^{\ast} c_m^{ } 
e^{i t (\epsilon_n - \epsilon_m)} \int_0^{2\pi} dx \
\langle \phi_n(t) | p_x | \phi_m(t)\rangle
\label{expand}
\end{equation}
where $c_n$ are expansion coefficients in the Floquet basis,
$\epsilon_n$ are the quasienergies,
$|\phi_m(t)\rangle$ are the Floquet states, and 
$p_x$ is the standard momentum operator.
It is important to note the off-diagonal interference
terms $\exp\left[i t (\epsilon_n - \epsilon_m)\right]$. 
If the system were strongly chaotic, level
repulsion would imply that the quasienergy separations are generally
large, and so these interferences would rapidly average to zero. This 
yields the approximate formula given in Eq. 1 of 
Ref. \onlinecite{liars}, in which solely the diagonal terms
of the current are retained, collapsing the coherent sum to
an incoherent one. This strong chaoticity would correspond to a short
Ehrenfest time, and so our analysis would similarly predict a short time 
scale for the stability of the ratchet current.

\begin{center}
\begin{figure}
\includegraphics[width=0.45\textwidth,clip=true]{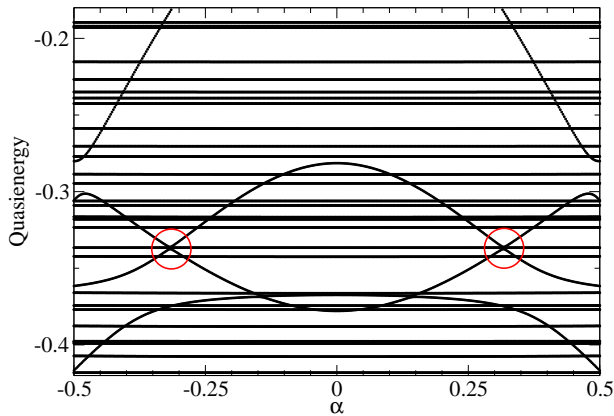}
\caption{Quasienergy spectrum of the ratchet system for time-symmetric
driving ($K=2.4$, $\omega=1$, $\beta=0$). The majority of the
quasienergies show little
dependence on the spatial asymmetry $\alpha$, but the narrow
avoided crossings (with a gap of $\Delta \epsilon \simeq 0.0014$)
at $\alpha = \pm 0.32$ (highlighted by red
circles) give rise to long-lived transient currents, with
a duration much longer than typical experimental observation times,
even though the temporal symmetry of the driving is not broken.}
\label{fig1}
\end{figure}
\end{center}

When the quasienergy spectrum contains degeneracies
the corresponding interference terms in Eq. \ref{expand} will not
decay (for exact degeneracies), or will only decay extremely
slowly (when the degeneracy is approximate). Although the analysis of Benenti
\etal \ cannot describe this situation, our approach would simply yield
a longer Ehrenfest time, indicating the enhanced stability 
of the current. Such a quasidegeneracy is actually present 
(see Fig. \ref{fig1}) in the numerical results
presented in the Comment. For a value of the asymmetry parameter
$\alpha=0.32$, a very narrow avoided crossing appears,
producing the long-lived current plotted in the inset
on Fig. 1 of  Ref. \onlinecite{liars}. The conclusion of Benenti \etal \
that ``no asymptotic directed transport occurs for any value of K'' is 
thus not generally correct -- it
depends on the detailed form of the quasienergy spectrum.

\begin{center}
\begin{figure}
\includegraphics[width=0.45\textwidth,clip=true]{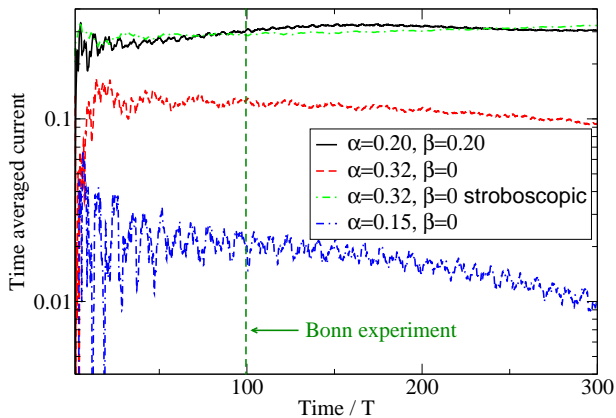}
\caption{Decay of the continuously averaged current $\langle I(t) \rangle$
for the strongly driven ratchet system, $K=2.4$, $\omega=1$,
for different asymmetry parameters $\alpha$ and $\beta$.
A typical experimental observation time, indicated by the vertical green line,
is taken from the recent work of the Bonn group \cite{bonn}.
For $\beta=0.2$ (black solid line) time-symmetry of
the driving is explicitly broken
and $\langle I(t) \rangle$ approaches a non-zero asymptotic value.
When $\beta=0$, however, the time-symmetry of the driving is not broken
and asymptotically the ratchet current must decay to zero.
For $\alpha=0.32$ (red dashed line) the average current
decays extremely slowly, due to a narrow avoided crossing
in the quasienergy spectrum (see Fig. \ref{fig1}).
We also plot the stroboscopically averaged value of the current
(green dot-dashed line)
for these driving parameters, which asymptotically approaches
a constant value.
For other values of $\alpha$ (blue dash-dot-dot line) the current decays
more rapidly, but nonetheless remains significant
over timescales much longer than those used in experiment.}
\label{fig2}
\end{figure}
\end{center}

Benenti \etal \ correctly note that ``the stroboscopic current...
remains finite forever''. We do not dispute this point, but it is irrelevant.
This would be an issue only if we had attempted to deduce the time scale
for the decay of the current by making a fit of the time-dependence of
the stroboscopically averaged current. As we emphasise above,
this was not our procedure.
Even making use of the
continuous time-average proposed by Benenti \etal, in place of
the more experimentally-relevant stroboscopic average
plotted in Fig. 3 of Ref. \onlinecite{prl}, the
conclusions of our Letter would be unaffected. In Fig. \ref{fig2} 
we show the decay rates of the continuously-averaged
current, which clearly show that even for time-symmetric driving,
significant ratchet currents are produced over timescales that
are very long in comparison to typical experimental 
observation times \cite{bonn}.
Although the interacting case ($g \neq 0$) is not amenable to  
Floquet analysis, very similar results are numerically obtained for the values 
of nonlinearity considered in Fig. 3 of Ref. \onlinecite{prl}.

Benenti \etal \ further attempt to support their case by considering the
behavior of the harmonic oscillator. This example is trivial; it is
not even periodically-driven. A more telling comparison would be with
the phenomenon of dynamical localization \cite{dyn_loc}. Here a 
particle on a lattice, subjected
to a driving potential, periodically expands and collapses when the parameters
of the driving are adjusted to certain specific ratios. Viewed
stroboscopically the particle appears to be frozen.
The purely stroboscopic character of this phenomenon does
not prevent it from being a genuine physical effect,
as reflected in the name ``dynamical localization''.

In summary, in our Letter we never claimed that for
time-symmetric driving a ratchet current would last forever (although
in the present Reply we point out that it could be possible if exact
quasienergy degeneracies existed). A stroboscopic simulation may indeed
overestimate the decay time of the ratchet current. However, our
decay estimate was based on general quantum
chaos theory arguments. Moreover, even a continuously time-averaged
current may exhibit ratchet behavior for times longer than present
experimental times. The conclusions of our Letter thus remain
unaffected.

\bigskip
C.E.~Creffield and F.~Sols \\

\noindent
Dpto de F\'isica de Materiales \\
Universidad Complutense de Madrid \\
E-28040 Madrid \\
Spain

\begin{thebibliography}{9}
\bibitem{liars}
{G.~Benenti, {\it et al.}, Phys. Rev. Lett. {\bf 104}, 228901 (2010).}

\bibitem{prl}
{C.E.~Creffield and F.~Sols, Phys. Rev. Lett. {\bf 103}, 200601 (2009).}

\bibitem{shep}
{J.~Martin, B.~Georgeot, and D L.~Shepelyansky, Phys. Rev. Lett.
{\bf 101}, 074102 (2008).}

\bibitem{bonn}
{T.~Salger, {\it et al.},
Science {\bf 326}, 1241 (2009).}

\bibitem{dyn_loc}
{ D H.~Dunlap and V.M.~Kenkre, Phys. Rev. B {\bf 34}, 3625 (1986).} 

\end{thebibliography}
\end{document}